\documentclass[twocolumn,showpacs,amsmath,amssymb,aps,prl,floatfix]{revtex4-1}
\usepackage[latin1]{inputenc}
\usepackage[american]{babel}
\usepackage[T1]{fontenc}
\usepackage{mathrsfs}
\usepackage{hyperref}
\usepackage{multirow}
\usepackage{bm}
\usepackage{epsfig}
\usepackage{graphicx}
\usepackage{subeqnarray}
\usepackage{bbold}
\usepackage{upgreek}
\usepackage{url}
\hypersetup{colorlinks=true, urlcolor=blue, citecolor=blue, filecolor=blue, linkcolor=blue}

\begin{document}

\title{Astrophysical line diagnosis requires non-linear dynamical atomic modeling}

\date{\today}

\author{Natalia~S.~Oreshkina}
\affiliation{Max-Planck-Institut f\"{u}r Kernphysik, Saupfercheckweg 1, 69117 Heidelberg, Germany}
\author{Stefano~M.~Cavaletto}
\affiliation{Max-Planck-Institut f\"{u}r Kernphysik, Saupfercheckweg 1, 69117 Heidelberg, Germany}
\author{Christoph~H.~Keitel}
\affiliation{Max-Planck-Institut f\"{u}r Kernphysik, Saupfercheckweg 1, 69117 Heidelberg, Germany}
\author{Zolt\'{a}n~Harman}
\affiliation{Max-Planck-Institut f\"{u}r Kernphysik, Saupfercheckweg 1, 69117 Heidelberg, Germany}

\begin{abstract}

Line intensities and oscillator strengths for the controversial 3C and 3D astrophysically relevant lines in
neonlike Fe${}^{16+}$ ions are calculated. We show that, for strong x-ray sources, the modeling of the spectral lines by a peak with an area proportional
to the oscillator strength is not sufficient and non-linear dynamical effects have to be taken into account.
Furthermore, a large-scale configuration-interaction calculation of oscillator strengths is performed with the inclusion of
higher-order electron-correlation effects. The dynamical effects give a possible resolution of discrepancies of theory and experiment
found by recent measurements, which motivates the use of light-matter interaction models also valid for strong light fields
in the analysis and interpretation of astrophysical and laboratory spectra.

\end{abstract}

%
%

\maketitle

Astrophysical spectra recorded by space observatories provide the only means to determine the element composition,
temperature, density, and velocity of distant celestial objects such as stars, x-ray binaries, black hole accretion discs, or active galactic
nuclei~\cite{Trigo2013,Liu2013,Bernitt2012,Behar2001,Xu2002,Paerels2003,Brown2001,vanParadijs1999,Rybicki2004}.
Such x-ray (or optical) spectra are often composed of a series of peaks associated with a range of elements, ionic
charge states, and transitions. Therefore, a large amount of reliable atomic data is needed to disentangle the physical
properties of the emitting objects. Such data---transition energies and probabilities, oscillator strengths, collisional and recombination
cross sections, etc.---may be obtained from laboratory astrophysics
experiments (see e.g. \cite{Bernitt2012,Hahn2014,Schnorr2013,Rudolph2013,Schmidt2007,Beiersdorfer2004,Beiersdorfer2001,Brown1998}) or, more
economically, from theoretical calculations (see e.g. \cite{Bhatia1992,Cornille1994,Safronova2001,Chen2001}).

The x-ray emission lines of highly charged Fe ions are among the brightest in astrophysical spectra. Within the last decade, several observations
were performed with the space laboratories Chandra and XMM-Newton (see e.g. \cite{Huenemoerder2011,Paerels2003,Xu2002,Mewe2001}).
The line-strength ratio of two $2p\to 3d$ lines in Fe${}^{16+}$, customarily denoted as 3C [$2p^6(J=0) \rightarrow (2p^5)_{1/2}3d_{3/2}~(J=1)$,
transition energy of 826~eV] and 3D [$2p^6(J=0) \rightarrow (2p^5)_{3/2}3d_{5/2}~(J=1)$, at 812~eV], was observed, but the results
disagreed with theoretical predictions~\cite{Bhatia1992,Cornille1994,Safronova2001,Chen2001}.
Initially this disagreement was considered to originate from the co-existence of different charge states of Fe,
and later, after laboratory measurements, as an effect of electron-impact excitation of the ion.
Furthermore, since several theoretical calculations of transition probabilities in highly charged ions agreed well with
the experiments (see, e.g.~\cite{Tupitsyn2005,Volotka2006,Lapierre2005}),
there was no reason to assume that essential contributions had not been included in the predictions.
The question was out of focus until the first laser spectroscopic experiment in the x-ray regime~\cite{Bernitt2012},
enabled by the advent of x-ray free-electron-laser (XFEL) facilities \cite{Emma2010}.
This experiment at the Linac Coherent Light Source (LCLS, Ref.~\cite{LCLS}) gave hints for an incorrect atomic structure theory:
a disagreement between all state-of-the-art theoretical predictions (ratio of the 3C and 3D oscillator strength around 3.5 and above)
and the experimental line-strength ratio of 2.61(23) has been stated~\cite{Bernitt2012}. In the comparison and in previous astrophysical modeling,
it was assumed that the intensity of a line is proportional to the electric dipole oscillator strength.

In this Letter, motivated by the above discrepancy, we refine the theory of x-ray--ion interactions by calculating higher-order electron-correlation and dynamical effects
contributing to the 3C/3D line-strength ratio. Our results suggest that the disagreement may be removed by the inclusion of non-linear dynamical effects
present in the case of strong driving x-ray fields. The broadening of the spectral lines due to the high x-ray intensity depends on the dipole moment of the
transitions involved, significantly influencing both the line strengths as well as their ratio.

{\it Higher-order correlation and QED corrections to the oscillator strengths.}
To improve the atomic structure theory of the dipole transition rates, we first apply a large-scale implementation
of the configuration-interaction Dirac-Fock-Sturm (CI-DFS) method~\cite{tup2003,Tupitsyn2005},
which uses radial wave functions obtained by the numerical solution
of the Dirac-Fock equation for the occupied orbitals ($1s$, $2s$, $2p$, $3d$), and virtual orbitals with positive and negative energies
represented by a relativistic Sturmian basis set.
These orbitals are employed to construct configuration state functions, i.e., Slater determinants
$| \Phi^i_J \rangle$ ($i=1,\dots ,N$, where $N$ is the total number of configurations) in an angular momentum-coupled basis
with a total angular momentum $J$. The atomic wave function
$| \Psi_J \rangle$ is finally represented as a linear combination of a large set of configurations:
\begin{eqnarray}
|\Psi_J \rangle = \sum_{i=1}^N c_i |\Phi_J^i \rangle .
\end{eqnarray}
The Einstein coefficients $A_{eg}$ and oscillator strengths $f_{eg}$ are calculated with such wave
functions representing the $2p^6$ ground $(g)$ and excited $(e \in \{(2p^5)_{1/2}3d_{3/2},(2p^5)_{3/2}3d_{5/2}\})$ states of the ion~\cite{Grant1974}:
\begin{eqnarray}
A_{eg} &=& \frac{4 \pi^2 e^2 c^2}{(2J_i+1)\omega_{eg}} \sum_{M_i,M_f}\sum_{\vec{k}/k, \sigma}
\left| \langle {e} | \vec{\alpha}\vec{\epsilon}_{\vec{k}\sigma}e^{-i\vec{k}\vec{r}}| {g} \rangle \right|^2\,, \nonumber \\
f_{eg} &=& \frac{2J_e+1}{2J_g+1}\frac{A_{eg}mc^3}{2\omega^2_{eg}e^2} \,.
\end{eqnarray}
Here, the summation goes over the magnetic quantum numbers of the initial and final states and the polarization $\sigma$ of the emitted photon, and, in addition,
an integration is performed over the direction $\vec{k}/k$ of the emitted photon. $c$, $e$ and $m$ denote here the speed of light, the elementary charge, and the electron mass,
respectively, and $\vec{\alpha}$ and $\vec{\epsilon}_{\vec{k}\sigma}$ stand for the vector of alpha matrices and the photon polarization unit vector.
The dimensionless oscillator strength is of particular interest, as for low driving-field intensities it is proportional to the line strength, here given as the
integrated peak area of the resonance photon scattering cross section~\cite{Grant1974,Foot}:
\begin{eqnarray}
S_{eg} = \frac{\pi^2c^2\hbar^3}{(\hbar\omega)^2}\frac{g_e}{g_g}A_{eg} \propto f_{eg}\,.
\end{eqnarray}
In order to match the accuracy of the experimental transition energies $\omega_{eg} = \omega_e-\omega_g$, additional quantum-electrodynamic (QED) corrections
are taken into account in an ab initio manner. The QED corrections in first order in the fine-structure constant $\alpha$ consist of the
self-energy (SE) and vacuum-polarization (VP) terms.
The SE correction is decomposed into zero-, one-,
and many-potential terms. The zero-potential and one-potential terms are calculated in momentum space
using formulas from Ref.~\cite{Yer1999}. The residual part of the SE correction, the so-called many-potential term, is
calculated in coordinate space.
For any given intermediate-state angular momenta, the summation over the Dirac spectrum
is performed using the dual-kinetic-balance approach~\cite{dkb} with basis functions constructed from B-splines.
The VP correction was calculated in the Uehling approximation~\cite{uehling}.
Electron-interaction contributions to the QED corrections were calculated by evaluating the single-electron QED diagrams
with an effective potential accounting for the screening of the remaining 9 electrons, as described in Ref.~\cite{Schnorr2013}.
We find that although screening effects significantly modify the single-electron QED corrections, the total QED effect on the
transition energies is on the 10-meV level.

While we can conclude that transition energies entering the theoretical oscillator strengths can be predicted with sufficient accuracy,
the discrepancy with the measurements prevails, and it may be rooted in the calculation of the
non-diagonal dipole matrix elements. In all previous theoretical studies it was assumed that the correlated many-electron wave function
can be well represented by constructing the configuration space with single and double electron exchanges from
the reference-state configuration. Here we also take into account
triple excitations, resulting in a huge number of configurations.
As a result, for the case of single and double excitations included, the 3C/3D oscillator strength ratio is 3.57, while the
contribution of the triple excitation is as low as -0.01. Thus, our results confirm earlier theoretical calculations and disprove
the significance of triple excitations, suggesting that the discrepancy of theory and experiment is not funded
in the inaccurate description of the ions' electronic structure.

{\it Modeling of strong-field effects.}
Once insufficiencies in structure calculations are ruled out, the next to investigate are dynamical aspects of light-matter interaction. In Ref.~\cite{Bernitt2012}
and in previous studies it was implicitly assumed that the intensity of the observed lines is proportional to the oscillator strength. This holds
true under the assumption of a relatively weak exciting field. However, nonlinearities are anticipated if the intensity $I$ of the field
is comparable to or larger than the saturation intensity $I_{\rm sat}$, to be defined below.
For the Fe transitions studied, with $I_{\rm sat} \approx$ $10^{11}$ W/cm$^2$, the intensity of LCLS pulses
is typically on or above this order of magnitude, such that the exciting field cannot be considered as weak anymore.

\begin{figure}[t]
\includegraphics[clip=true, width=1.0 \columnwidth]{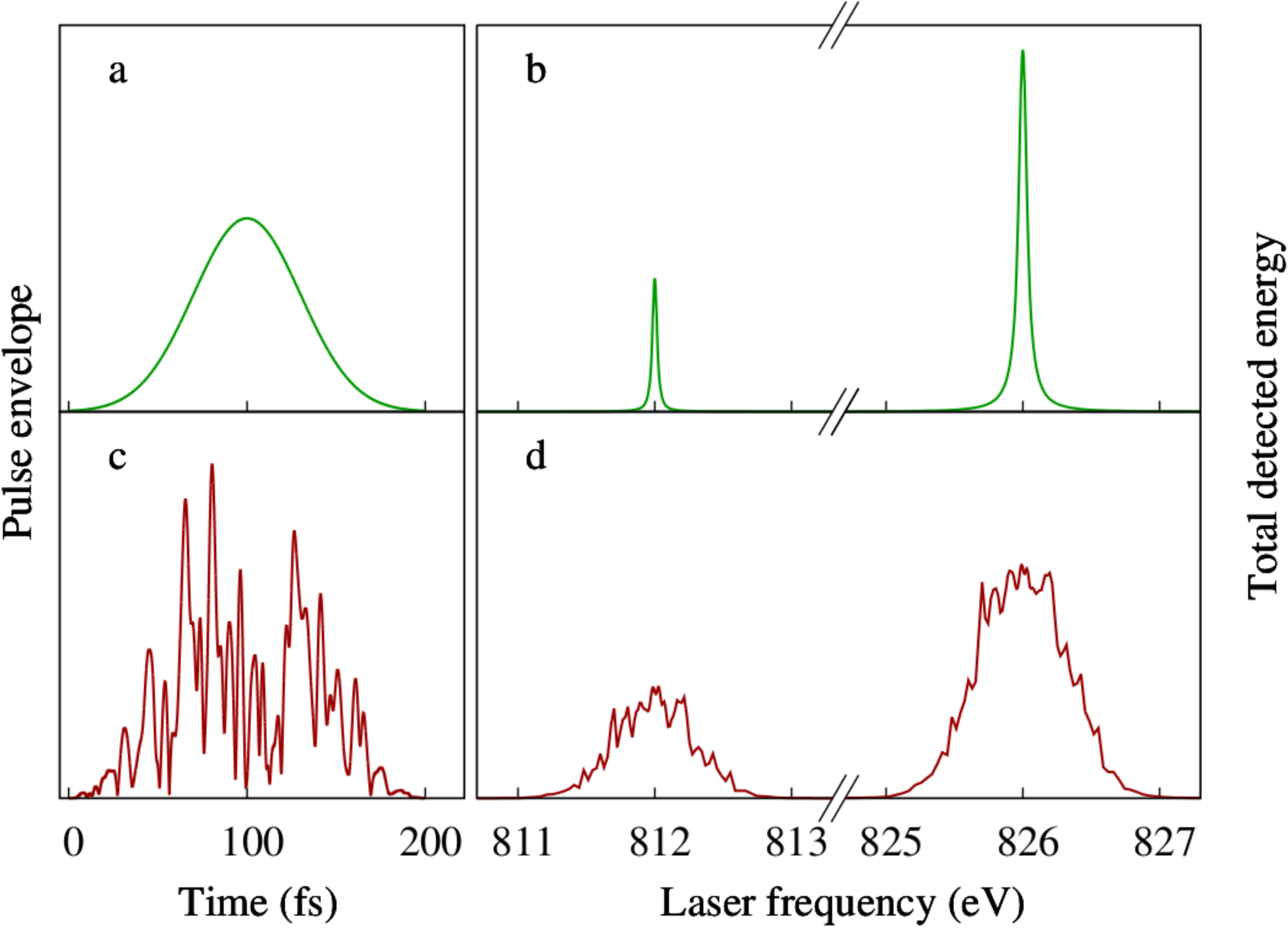}
\caption{(a) Time-envelope of a Gaussian and (c) an incoherent pulse, in arbitrary units.
(b,d) The fluorescence signal (total energy of fluorescence photons) as a function of the XFEL photon energy
in the energy range of the 3C and 3D transitions in Fe${}^{16+}$, in arbitrary units. Here, an x-ray intensity of $I=10^{13}$~W/cm$^2$ was used and results
are shown for (c) a Gaussian pulse and (d) after averaging over 10 incoherent pulses with a time duration of $T=200$~fs. For incoherent pulses,
the bandwidth is set to $B=1$~eV.}
\label{fig:spectrum}
\end{figure}

We therefore improve the physical description and perform time-dependent simulations by modeling the ion as a two-level
system with ground state $|g\rangle$ and excited state $|e\rangle$.
The transition energy $\omega_{eg}$ approximately equals 826~eV and 812~eV for the 3C and 3D lines, respectively,
and the decay width is $\Gamma_{eg} = \hbar A_{eg}$.
We describe the atomic system via the density matrix $\hat{\rho}(t)$ of elements $\rho_{ij}$, with $i,\,j\in\{g,\,e\}$,
whose evolution in time is given by the master equation \cite{Scully,Kiffner2010}
\begin{eqnarray}
\frac{d\hat{\rho}}{dt}= - \frac{i}{\hbar}[\hat{H},\hat{\rho}(t)] + L\hat{\rho}(t)\,,
\end{eqnarray}
where the square brackets stand for a commutator, and the Lindblad superoperator $L$ represents the spontaneous
decay from the exited state $|e\rangle$ to the ground state $|g\rangle$
with decay rates equal to $2.22\times 10^{13}$~s$^{-1}$ and $6.02\times 10^{12}$~s$^{-1}$ for the 3C and the 3D 
transitions, respectively, according to our CI-DFS calculations. The Hamiltonian $\hat{H} = \hat{H}_0 + \hat{H}_{\mathrm{int}}$ is the sum of
the electronic-structure Hamiltonian $\hat{H}_0 = \sum_{i \in \{g,e\}}\hbar\omega_i |i\rangle\langle i|$ and of the
Hamiltonian $\hat{H}_{\mathrm{int}}$ describing the interaction of the ion with an external time-dependent 
electric field $\mathcal{E}(t) = E(t)\,\cos(\omega_{\mathrm{X}}t+\psi(t))$ of x-ray carrier frequency
$\omega_{\mathrm{X}}$, envelope $E(t)$, and phase $\psi(t)$. We introduce the vector $\vec{R}(t)$ of the slowly varying components of the density matrix,
\begin{equation}
\vec{R}(t) = (\rho_{gg}(t),\,\rho_{ge}(t)\,e^{-i\omega_{\mathrm{X}}t},\,\rho_{eg}(t)\,e^{i\omega_{\mathrm{X}}t},\,\rho_{ee}(t))^{\mathrm{T}}\,,
\end{equation}
such that the master equation can be written in the matrix form
\begin{equation}
\frac{d\vec{R}(t)}{dt} = {\bf M}(t)\vec{R}(t)\,,
\label{eq:master-equation}
\end{equation}
with the $4\times 4$ time-dependent matrix
\begin{eqnarray*}
{\bf M}(t) = \left(
\begin{array}{c c c c}
0 				& -i\frac{\Omega^*_R(t)}{2}	& i\frac{\Omega_R(t)}{2}	&  \Gamma_{eg}			\\
-i\frac{\Omega_R(t)}{2}		& i\Delta-\frac{\Gamma_{eg}}{2}	& 0 				&  i\frac{\Omega_R(t)}{2}	\\
 i\frac{\Omega^*_R(t)}{2}	& 0 				& -i\Delta-\frac{\Gamma_{eg}}{2}& -i\frac{\Omega^*_R(t)}{2}	\\
0 				& i\frac{\Omega^*_R(t)}{2}	& -i\frac{\Omega_R(t)}{2}	& -\Gamma_{eg}			\\
\end{array} \right)\,.
\end{eqnarray*}
The complex time-dependent Rabi frequency $\Omega_R(t)=eE(t)\langle g| \hat{r} | e \rangle e^{i\psi(t)}/\hbar$ is proportional to the square root of the x-ray intensity $I$
and $\Delta=\omega_{eg}-\omega_X$ denotes the detuning of the laser frequency from the transition frequency. For a continuous-wave driving field,
$E(t) = \bar{E}$, $\psi(t) =0$, with correspondingly constant Rabi frequency $\bar{\Omega}_R$, the solution of the master equation~(\ref{eq:master-equation}) converges
for $t\rightarrow \infty$ to the stationary solution \cite{Scully,Foot}
\begin{equation}
\bar{R}_{ee}(\Delta) = \frac{\bar{\Omega}_R^2}{4\Delta^2 + \Gamma_{eg}^2 + 2\bar{\Omega}_R^2}\,.
\end{equation}
The energy detected per unit time for a given detuning $\Delta$ is
\begin{equation}
\mathcal{I}(\Delta) \propto \Gamma_{eg}\omega_{eg}\bar{R}_{ee}(\Delta)\,,
\end{equation}
which can be used to calculate the ratio of the intensities emitted by the two lines by means of the integrals over the detuning $\Delta$
\begin{equation}
S_{\rm 3C}/S_{\rm 3D} = \frac{\int {\rm d}\Delta \mathcal{I}_{\rm 3C}(\Delta)}{\int {\rm d}\Delta \mathcal{I}_{\rm 3D}(\Delta)} \,.
\label{eq:analyticalratio}
\end{equation}
By introducing for each line the saturation intensity $I_{\mathrm{sat}} = I \,\Gamma_{eg}^2/(2 \bar{\Omega}^2_R)$, this leads to
\begin{equation}
S_{\rm 3C}/S_{\rm 3D} = \frac{\Gamma_{\mathrm{3C}}\omega_{\mathrm{3D}}^2}{\Gamma_{\mathrm{3D}}\omega_{\mathrm{3C}}^2}\,\sqrt{\frac{1 + I/I_{\rm sat,3D}}{1+ I/I_{\rm sat, 3C}}}.\label{eq:int-ratio}
\end{equation}
For a weak exciting field ($I\ll I_{\rm sat}$), this agrees with the linear theory of resonance fluorescence predicting a ratio of 3.56, while in the strong-field limit
($I\gg I_{\rm sat}$) Eq.~(\ref{eq:int-ratio}) converges to the value $S_{\rm 3C}/S_{\rm 3D} \rightarrow 7.03$.

\begin{figure}[t]
\includegraphics[clip=true, width=1.0 \columnwidth]{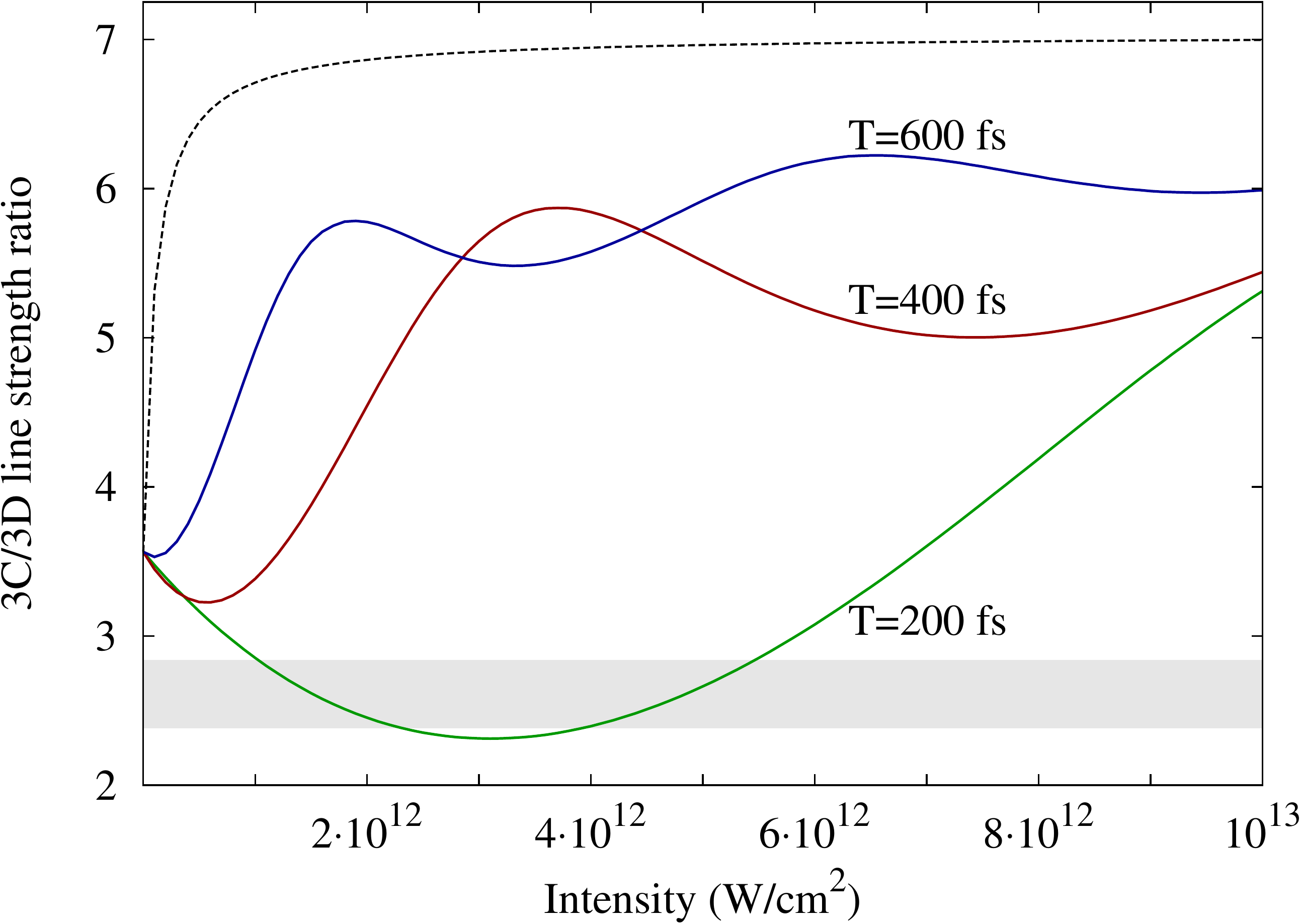}
\caption{The line-strength ratio
$S_{\rm 3C}/S_{\rm 3D}$ as a function of the intensity and duration of the Gaussian pulse.
The dashed line is for a continuous wave field (Eq.~\ref{eq:analyticalratio}).
The gray shaded area shows the experimental ratio 2.61(23)~\cite{Bernitt2012} together with its error bar.}
\label{fig:function}
\end{figure}

For a time-dependent (pulsed) driving field, the system of differential equations~(\ref{eq:master-equation}) is solved assuming the initial conditions
$\vec{R}_0=(1,0,0,0)^T$. Thereby, we can calculate the total detected energy as a function of the detuning,
\begin{equation}
\mathcal{E}(\Delta) \propto \Gamma_R\omega_{eg}\int_{-\infty}^{+\infty}R_{ee}(t)dt\,,
\end{equation}
where $R_{ee}(t)$ also depends on $\Delta$. This yields a spectrum which can be compared to the experimentally measured one.
The ratio of total emitted line energies can be again calculated as a ratio of the integrals over the detuning $\Delta$:
\begin{equation}\label{eq:spectrum}
S_{\rm 3C}/S_{\rm 3D} = \frac{\int {\rm d}\Delta \mathcal{E}_{\rm 3C}(\Delta)}{\int {\rm d}\Delta \mathcal{E}_{\rm 3D}(\Delta)} \,.
\end{equation}
Typical LCLS intensities are in the range of $I=10^{11}-10^{14}$~W/cm$^2$ and pulse durations in the range $T=200-2000$~fs.
Also, not the total electromagnetic energy emitted in all directions was detected in the experiment~\cite{Bernitt2012}, but only a fraction of it emitted into a given solid angle.
However, since the excited states have the same symmetry (total angular momentum $J=1$) and the ground state of the ion is spherically symmetric ($J=0$), the angular
distribution of the emitted radiation is the same for both transitions, allowing one to compare the strength ratio as defined above with the experimentally determined line
intensities.

{\it Matching the experimental line strengths and discussion.}
In Fig.~\ref{fig:spectrum}b, simulated 3C and 3D fluorescence lines are presented assuming strong XFEL pulses of Gaussian shape for an
intensity of $I=10^{13}$~W/cm${}^2$ and duration of $T=200$~fs. We use the pulse envelope
$E(t) = E_{\rm max} e^{- \frac{t^2}{T^2} 32 \ln(2)}$ and a constant phase $\psi(t)=0$.
The ratio of the 3C and 3D line strengths is shown separately on Fig.~\ref{fig:function} as a function of pulse parameters.
The strengths and their ratio are sensitive to the change of pulse intensity and duration. Between $I=1-6\times 10^{12}$ W/cm$^2$ and for
$T=200$~fs, the resulting line-strength ratio is in the range of $2.31-3.08$, as presented in Fig.~\ref{fig:function}. This result is still in
agreement with the measured value of 2.61$\pm$0.23. We also infer that the pulse-by-pulse variation of intensity and duration contributes to the
experimental uncertainty of the ratio.
These results confirm the importance of strong-field dynamical effects in a relatively intense XFEL field. We note that, in this range of parameters,
we are still in the weakly nonlinear regime: below intensities of approximately 10$^{11}$~W/cm$^2$, the linear theory of resonance fluorescence may be applied.
(The linear model is also applicable for transitions with significantly lower dipole matrix elements, even if the intensity of the x-ray source is high.)
At higher intensities, or for longer pulses, which is also still in the range of possible experimental parameters, the sensitivity to the pulse parameters increases,
resulting in an oscillation of the line-strength ratio of the 3C and 3D lines between 5.5 and 6.5, in agreement with the previously mentioned
intensity-saturation effects. Increasing the pulse duration in the simulations, one reaches the limit of continuous-wave fields, which is shown by the dashed line
[cf. Eq.~(\ref{eq:analyticalratio})].

\begin{figure}[t]
\includegraphics[clip=true, width=1.0 \columnwidth]{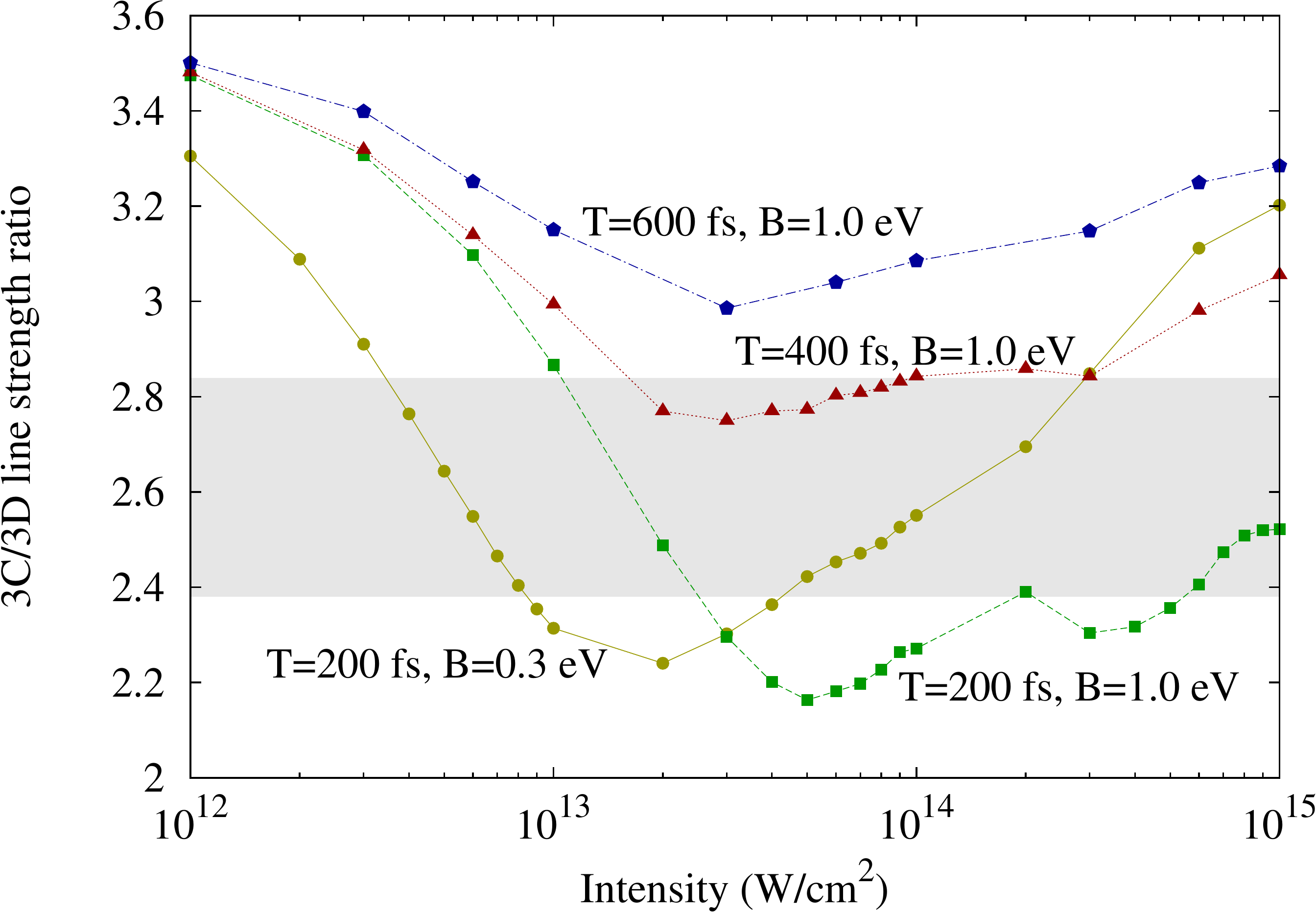}
\caption{The line-strength ratio
$S_{\rm 3C}/S_{\rm 3D}$ as a function of the intensity $I$ and duration $T$ of the incoherent pulse.
For the shortest pulses, results with two different bandwidths $B$ are shown.
The gray shaded area shows the experimental ratio 2.61(23)~\cite{Bernitt2012} together with its error bar.}
\label{fig:avpulse}
\end{figure}

For an even more realistic modeling, we additionally take into account the chaotic nature of XFEL pulses generated via self-amplified
spontaneous emission \cite{Bonifacio1984}, by modeling amplitude $E(t)$ and phase $\psi(t)$  via
the partial-coherence method from Refs.~\cite{Pfeifer2010,Cavaletto2012}. An understanding of incoherence effects is not only relevant for laboratory measurements but
also for astrophysical observations, as natural x-ray sources lack coherence. In the simulations, we employ a series of such randomized pulses, an example of which
is shown on Fig.~\ref{fig:spectrum}c. Fig. \ref{fig:spectrum}d presents the fluorescence signal resulting from the use of such incoherent pulses, while Fig. \ref{fig:avpulse}
displays the line-strength ratio \eqref{eq:spectrum} obtained with chaotic pulses of different duration, intensity and bandwidth.
Each point in Fig.~\ref{fig:avpulse} is obtained by averaging over 10 independent realizations of a chaotic pulse.
The numerical uncertainty of the obtained results is estimated to be on the level of $1-2\%$. As in the case of Gaussian pulses, the line-strength ratio
clearly depends on the pulse parameters. Also here, for small intensities, the value of 3.56 is approached,
in agreement with the linear theory of resonance fluorescence. We notice that, for
incoherent pulses, the effect of the decrease in the 3C/3D line-strength ratio can be observed within
a wider range of pulse intensities than for Gaussian pulses, as well as for a significantly larger interval of pulse durations.
As shown in the picture, this effect becomes more significant at increasing values of the bandwidth (i.e., the FWHM of the energy spectrum)
of the chaotic pulse, which in the experiment~\cite{Bernitt2012} can be estimated to be of the order of 1~eV.
The decrease in bandwidth (increase in coherence time) corresponds to a more coherent pulse, and a behaviour closer to that displayed
by fully coherent (transform-limited) Gaussian pulses.

Although parameters of XFEL pulses are not fixed from pulse to pulse because of their chaotic nature,
the average peak intensity may be estimated to lie in the considered range. However, the predicted decrease
in the 3C/3D ratio within a broad interval of pulse intensities and pulse durations allow us to suggest
that the observed unexpectedly low value of the 3C/3D ratio is determined by previously neglected nonlinear dynamical effects.

In summary, we conclude that a new approach is called for in astrophysical line diagnostics, taking into account effects depending on the intensity
of the radiation field. Above certain intensities, weak-field atomic theory may not give a proper picture of radiative processes
taking place in, e.g., black hole accretion discs. For stellar-mass black holes, the Eddington luminosity of approx. $10^{38}$~erg/s translates to a total radiation intensity
(integrated over the Planck distribution at a temperature of $5 \times 10^7$~K~\cite{Shapiro1983}) on the order of $10^{18}$~W/cm$^2$ at a distance of 3 Schwarzschild radii,
of which approximately $3 \times 10^{13}$~W/cm$^2$ is in a typical 1-eV range of an atomic line at 1~keV. Furthermore, such nonlinear dynamic effects also have to be considered
in laboratory astrophysics experiments employing sources of high x-ray intensities, such as, e.g., x-ray free-electron lasers.

We acknowledge helpful advice from Ilya I. Tupitsyn and insightful conversations with Alberto Benedetti.


%

\end{document}